\documentstyle[12pt]{article}
\begin{document}
\thispagestyle{empty}
\begin{center}
{\LARGE \tt \bf On dilaton solutions of de Sitter inflation and primordial spin-torsion density fluctuations}
\end{center}
\vspace{0.5cm}
\begin{center}
{\large \tt \bf L.C. Garcia de Andrade\footnote{Departamento de F\'{\i}sica 
Teorica, Instituto de F\'{\i}sica,Universidade do Estado do Rio de Janeiro,Rua Sao Francisco Xavier 524,Maracana,UERJ,Brasil-CEP:20550-003}}
\end{center}
\vspace{2cm}
\begin{abstract}
Four classes of exact solutions of Einstein-Cartan dilatonic inflationary de Sitter cosmology are given.The first is obtained from the equation of state of massless dilaton instead of an unpolarized fermion fluid used previously by Gasperini.Repulsive gravity is found in the case where dilatons are constraint by the presence of spin-torsion effects.The second and third solutions represent respectively massive dilatons in the radiation era with the massive potential and torsion kinks and finally the dust of spinning particles.Primordial spin-density fluctuations are also computed based on Primordial fluctuations of temperature obtained from COBE data.The temperature fluctuation can also be computed from the nearly flat spectrum of the gravitational waves produced during inflation and by the result that the dilaton mass would be proportional to the Hubble constant.This result agrees with the COBE data.This idea is also used to compute the spin-torsion density in the inflation era.
\end{abstract}
\noindent
\section{Introduction}
Dilatonic Cosmological models have been recently investigated in great detail in Cosmology \cite{1} and Supergravity \cite{2}.Earlier the role of spin-torsion as a source of repulsive gravity have been investigated in detail by Gasperini \cite{3} and Cosimo Stornaiolo in his PhD.thesis \cite{4}.More recently we have investigate the role of torsion on inflating defects \cite{5}.Also recently Maroto and Shapiro have investigated a cosmological model of de Sitter type with dilatons and torsion in Higher-order gravity \cite{6}.In this letter we solve the Einstein-Cartan dilaton cosmology equations of gravity for the case of a de Sitter model where the equation of state given a priori.The first is a massless dilatonic case.The three others deal with massive dilatons in three distinct cases.The first and second is the radiation era case for torsion kinks and massive potential.The last one handles with the dust of spinning particles with Cartan torsion \cite{7}.The temperature fluctuation is computed by using the nearly flat spectrum produced during inflation by gravitational waves and the solution obtained in our radiation era model that shows that the dilaton mass equals the Hubble constant.The resul is compared with COBE data.The spin-torsion fluctuation is also computed in a similar way.
\section{Spin-torsion effects in Dilatonic Inflationary Cosmology}
Let us now consider the cosmological metric for the spatially flat section to be given by 
\begin{equation}
ds^{2}=dt^{2}-a^{2}(dx^{2}+dy^{2}+dz^{2}) 
\label{1}
\end{equation}
The Einstein-Cartan equations are given by \cite{9,10} 
\begin{equation}
H^{2}=\frac{8{\pi}G}{3}({\rho}_{eff}-2{\pi}G{\sigma}^{2})
\label{2}
\end{equation}
and
\begin{equation}
\dot{H}+H^{2}=-\frac{4{\pi}G}{3}({\rho}_{eff}+3p_{eff}-8{\pi}G{\sigma}^{2})
\label{3}
\end{equation}
These two equations combined produced the following conservation equation 
\begin{equation}
(\dot{{\rho}_{eff}}-2{\pi}G(\dot{{\sigma}^{2}}))+3H({\rho}_{eff}+p_{eff}-4{\pi}G{\sigma}^{2})=0
\label{4}
\end{equation}
here $H=\frac{\dot{a}}{a}$ ,and ${\sigma}^{2}=<S_{ijk}S^{ijk}>$ where $S_{ijk}$,and $(i,j=0,1,2,3)$,represents the spin tensor and the spins are considered ramdomly oriented since at the very early Universe due to the very high temperatures and broken symmetries is not possible to have polarized matter and ${\rho}_{eff}$,and $p_{eff}$ are given respectively by 
\begin{equation}
{\rho}_{eff}={\dot{\phi}}^{2}+V({\phi})
\label{5}
\end{equation}
and
\begin{equation}
p_{eff}={\dot{\phi}}^{2}-V({\phi})
\label{6}
\end{equation}
we notice that by imposing the "stiff" matter equation of state $p_{eff}={\rho}_{eff}$ where ${\phi}$ is the dilaton potential and $V$ is the dilaton energy potential,implies that $V({\phi})$ vanishes, which represents massless dilatons.Now let us equate equations (\ref{3}) and (\ref{4}),this yields
\begin{equation}
{\dot{\phi}}^{2}={\pi}G{\sigma}^{2}
\label{7}
\end{equation}
Use of the conservation equation yields
\begin{equation}
{\dot{\phi}}^{2}=2{\pi}G{\sigma}^{2}+e^{-6Ht}
\label{8}
\end{equation}
Comparison of these two last equations yields an expression for the spin-torsion density in terms of time
\begin{equation}
{\sigma}^{2}=\frac{1}{{\pi}G}e^{-6Ht}
\label{9}
\end{equation}
Substitution of (\ref{9}) into the expression for the Hawking-Penrose convergence condition
\begin{equation}
R_{ik}u^{i}u^{k}=-3(H^{2})=16{\pi}G({\dot{\phi}}^{2}-2{\pi}G{\sigma}^{2})=-16{\pi}G{\sigma}^{2}<0
\label{10}
\end{equation}
shows that it is violated and a repulsive gravity is obtained.Cosmological perturbation densities can then be computed and the spin-torsion density determined.The computation without dilatons have been recently addressed by D.Palle \cite{8,9}.In the next section we perform these computations for the cases of radiation and hadron eras.
\section{De Sitter Inflation in the Radiation Era and spin-torsion density primordial fluctuations}
In this section we shall consider the equation of state for the radiation era $p_{eff}=\frac{1}{3}{\rho}_{eff}$.This condition yields
\begin{equation}
-{\dot{\phi}}^{2}+2V({\phi})=0
\label{11}
\end{equation}
This equation is interesting because can be solved for 
different potentials like torsion kinks and for the mass 
potential $V({\phi})=\frac{m^{2}}{2}{\phi}^{2}$ for the 
massive dilaton.By substituting this last expression into 
equation (\ref{11})
yields the following dilaton field equation
\begin{equation}
\dot{\phi}=-m{\phi}
\label{12}
\end{equation}
where we choose the minus sign for later convenience.The dilaton solution is 
\begin{equation}
{\phi}(t)=ae^{-mt}
\label{13}
\end{equation}
Nevertheless  equation (\ref{11}) should be matched with the well known dilaton equation
\begin{equation}
{\ddot{\phi}}+3H{\dot{\phi}}=-\frac{{\partial}V({\phi})}{{\partial}{\phi}}
\label{14}
\end{equation}
which solution is
\begin{equation}
{\phi}=e^{-\frac{3H}{2}t}
\label{15}
\end{equation}
The spin-torsion density does not appear in this equation since it does not interact with the dilaton field.With this matching we obtain $m=\frac{3H}{2}$ and the constant $a=1$.Therefore dilaton mass can be expressed in terms of the Hubble constant.
Equating equations (\ref{2}) and (\ref{3}) yields
\begin{equation}
2({\rho}_{eff}-2{\pi}G{\sigma}^{2})=-({\rho}_{eff}+3p_{eff}-8{\pi}G{\sigma}^{2})
\label{16}
\end{equation} 
and substitution of the radiation equation of state into equation (\ref{14}) yields
\begin{equation}
{\rho}_{eff}=3{\pi}G{\sigma}^{2}
\label{17}
\end{equation}
Substitution of (\ref{15}) into equation (\ref{4}) yields
\begin{equation}
\dot{{\sigma}^{2}}+8H{\sigma}^{2}=0
\label{18}
\end{equation}
which yields 
\begin{equation}
{\sigma}^{2}=e^{-8Ht}
\label{19}
\end{equation}
and ${\rho}_{eff}=3V({\phi})$.From (\ref{15}) yields 
\begin{equation}
V({\phi})={\pi}G{\sigma}
\label{20}
\end{equation}
From expression (\ref{18}) and (\ref{19}) we note that the 
dilaton vanishes at recent epochs of the Universe as well as 
the spin-torsion density and no spin-torsion dominated 
inflation  over the dilatonic inflation exists.Also this 
fact is also in agreement with the dificulty in observing 
torsion effects on cosmology from the classical point of view.
Similar aspects were discussed recently by Maroto and Shapiro \cite{6}.
Note that the dust case is very similar since equation 
(\ref{11}) is the same except that the factor of two in front 
of the potential should be dropped.Let us now examine that 
radiation era for a torsion kink potential $V({\phi})={\alpha}
{\phi}^{4}$.In this case equation (\ref{11}) changes to 
$-{\dot{\phi}}^{2}+{\alpha}{\phi}^{4}=0$ which has a simple 
solution ${\phi}=({\alpha}^{\frac{1}{2}}t)^{-1}$.We would 
like to mention that also in the radiation era the repulsive 
gravity is in effect since the violation of the geodesic 
convergence condition is maintained in this case.From the COBE data we know that $\frac{{\delta}T}{T}=10^{-5}$.The spectrum of gravitational waves with a temperature fluctuation of the order $\frac{{\delta}T}{T}=\frac{H}{m_{Pl}}$ and in our model there is a relation between the dilatonic mass and the Hubble constant thus $H=\frac{2}{3}m=10^{-4}m_{Pl}$ the temperature fluctuation yields the value $\frac{{\delta}T}{T}=10^{-4}$ which agrees with the value observed by COBE.From formula (\ref{19}) and the density of temperature based on the gravitational wave spectrum one must infer that $H=\frac{ln{{\sigma}^{2}}}{8t}$.Substitution of this value into the gravitational wave spectrum density we obtain $\frac{{\delta}T}{T}=\frac{ln{{\sigma}^{2}}}{8m_{Pl}t}=10^{-5}$ by the COBE data ,this in turn implies that ${\sigma}^{2}=e^{-8.10^{-5t}}$.At the inflation era $(t=10^{-35s})$ the value of the spin-torsion density becomes
${\sigma}^{2}=e^{-10^{-45}s}$.This is an extremely big value if we think that today's Universe possess a spin-torsion density of $e^{-10^{15 billion years}}$!.
\section*{Acknowledgements}
I am very much indebt to Prof.M.Gasperini,Prof.Shapiro,Prof.R.Ramos and Prof.H.P. de Oliveira for helpful discussions on the subject of this paper. Financial support from CNPq. and UERJ (FAPERJ) is gratefully acknowledged.


\begin{thebibliography}{10}
\bibitem{1}A.Vilenkin and P.S.Shellard,Cosmic Strings and other Topological Defects,Cambridge University Press(1995).
\bibitem{2}M.Cvetic and H.Soleng,Supergravity and domain wall,Physics Report(1997).
\bibitem{3}M.Gasperini,Repulsive gravity in the Very early Universe,gr-qc/9805060. 
\bibitem{4}C.Stornaiolo,Cosmology in Einstein-Cartan-Kibble-Sciama,DSc.thesis Naples University,(1985).
\bibitem{5}L.C.Garcia de Andrade,Inflating torsion defects and 
dilatonic domain walls,Class. and Quantum Gravity,(1999),June issue in press.
\bibitem{6}A.L.Maroto and I.L.Shapiro,Phys.Lett.B,414(1997)34.
\bibitem{7}V.de Sabbata and C.Sivaram,Spin and Torsion in Gravity,(1997).
\bibitem{8}M.Gasperini,Phys.Rev.Lett.569,(1986),2873.
\bibitem{9}D.Palle,On Primordial Cosmological Density Fluctuations in the Einstein-Cartan Gravity from COBE Data,gr-qc,(1999),Los Alamos Archives.
\bibitem{10}D.Palle,On Certain Relationships Between Cosmological Observables in the Einstein-Cartan Gravity,gr-qc,Los Alamos Electronic Archives,(1999).
\end{thebibliography}
\end{document}